\documentclass{ws-p8-50x6-00}

\begin{document}

\title{Interferometry in astrophysics as a roadmap for interferometry
in multiparticle dynamics}

\author{L. I. Gurvits}

\address{Joint Institute for VLBI in Europe\\  
P.O.Box 2, 7990 AA Dwingeloo, The Netherlands\\ 
E-mail: lgurvits@jive.nl}

\maketitle

\abstracts{
Interferometry is one of the most powerful experimental tools of modern
astrophysics. Some of its methods are considered in view of potential
applicability to studies of correlations in multiparticle dynamics.
}

\section{Introduction}

The reason for presenting interferometry as a tool of modern
astrophysics at a symposium on multiparticle dynamics arises from the
similarity of terms and methods used in both disciplines. It is
impossible to cover the whole subject of astrophysical interferometry
in one brief presentation. By necessity, it will be only a glimpse of
the subject; readers interested in a deep and comprehensive view on the
astrophysical interferometry are referred to the monograph by
Thompson et al.\cite{tms}. A historical overview of the development of
interferometry in radio astronomy has been given recently by Kellermann
\& Moran\cite{kik-jmm}.

The method descends from the seminal work by
Michelson\cite{mich90}. Using a two-slit optical
interferometer he succeeded in measuring the diameter of stars. Over
the following century, the method broadened its application in the
wavelength domain from the ultraviolet ($\lambda \sim 10^{-7}$~m) to
decametric radio waves ($\lambda \sim 10$~m) and in baseline length
from several meters to $\sim 10^{4}$~km.

\begin{figure}
\centerline{
\psfig{file=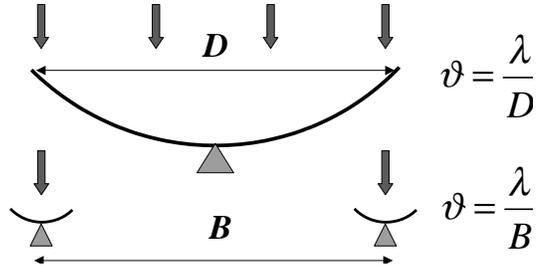,angle=-90,width=7cm,bbllx=203pt,bblly=196pt,bburx=421pt,bbury=629pt,clip=}
}
\caption{A telescope with aperture diameter $D$ observing at wavelength
$\lambda$ has diffraction-limited angular resolution $\theta \simeq
\lambda/D$ (upper panel). At the same frequency, a two-element
interferometer has an angular resolution along the direction of its
baseline $B$, of $\theta \simeq \lambda/B$ (lower panel), equal to the
angular resolution of the ``single-dish'' telescope if $B=D$. Incoming
radiation is shown by arrows.}
\label{fig:lambda-b}
\end{figure}

The main motivation for exploiting interferometry in a variety of
astronomical applications is illustrated in Fig.~\ref{fig:lambda-b}. To
achieve the desired angular resolution at a particular wavelength
$\lambda$ one has to place detectors of the emission at a particular
distance dictated by the diffraction. This distance could be either a
``diameter'' of a conventional (``single-dish'' in radio astronomical
slang) telescope, $D$, or a baseline of a two-element interferometer,
$B$. In many cases, technological complexity and cost of the former
option are prohibitively high; the only remaining possibility is to
create an interferometer. Of course, it comes at a price: the
sensitivity of the system is roughly proportional to its collecting
area, which is much smaller for the interferometer comparing to a
``full aperture'' telescope of the same angular resolution.

An important generalization of the interferometric technique is the
method of aperture synthesis. Its essence is in combining as many
interferometric pairs as possible, each of different length and/or
orientation. A certain calculation involving the responses of each pair
makes it possible to reconstruct (or, rather, to approximate) a
response of a telescope with a filled aperture. The method of aperture
synthesis was pioneered for radio astronomical applications in 1950's
in Cambridge by Ryle \& Hewish\cite{ry-hew} and their students and in
Australia by Christansen \& Warburton\cite{chr-war}.

At present, the interferometry and aperture synthesis are amalgamated
in one of the most productive astronomical techniques. Arguably, its
highest achievements so far are in the radio domain where modern
technology enables to operate Very Long Baseline Interferometers (VLBI)
with baselines comparable to an Earth diameter (see review by
Kellermann \& Moran\cite{kik-jmm}) and even exceeding the size of our
planet by placing one radio telescope in space (Hirabayashi et
al.\cite{hirax}).  VLBI routinely achieves milli-arcsecond ($1\,{\rm
mas} \approx 4.8\times 10^{-9}$~radian) and sub-milliarcsecond angular
resolution and is moving toward the micro-arcsecond angular scale by
observing at frequencies of tens and hundreds GHz (mm radio waves).
Optical interferometry is heading toward a milliarcsecond-scale imaging
with baselines of hundreds of meters (Lena \&
Quirrenbach\cite{len-qui}).

\section{ How it works: a simplistic view }

The output signal of a simple single-baseline (two-element) broad-band
interferometer observing a point-like source could be presented
as\cite{tms}
\be
r(\tau)=\lim_{T\rightarrow \infty}\frac{1}{2T}\int_{-T}^{T}V(t)V(t-\tau)\,dt,
\label{eq:corr-coef}
\ee
where $V$ is the voltage delivered by each element in response to the
source radiation, $2T$ is the integration interval, $t$ is the time, and
$\tau$ is the time delay between the signal detections by the two
elements.  Obviously, $r(\tau)$ is an unnormalized autocorrelation
function. The power spectrum of a signal is the Fourier transform
(shown below by the ``bi-directional harpoon'') of the autocorrelation
function of that signal, thus
\be
V_{1T}\star V_{2T} \rightleftharpoons \hat{V}_{1T}\hat{V}_{2T},
\label{eq:wie-khi}
\ee
where the left side represents the cross-correlation of
the two voltages at the telescopes, ``1'' and ``2'', while the right
side is the cross power spectrum of the signals $V_{1T}(t)$ and
$V_{2T}(t)$, represented by their Fourier images $\hat{V}_{1T}$ and
$\hat{V}_{2T}$, respectively. The usefulness of the
relation~(\ref{eq:wie-khi}) is in the possibility to estimate the cross
power spectrum by measuring the output signal of the correlator. The
equation~(\ref{eq:wie-khi}) is a generalization of the Wiener--Khinchin
relation.

In most astrophysical experiments the goal is to obtain a source
brightness distribution, $I(x,y)$, -- a function which formally
represents an image of the source in any arbitrary coordinate frame
$(x,y)$ (e.g. Cartesian coordinates in the plane tangent to the
celestial sphere at the position of the source). An ideal aperture
synthesis system measures the mutual coherence function $\Gamma (u,v)$
which is related to the brightness distribution of the source as
follows:
\be
\Gamma (u,v) = \int_{-\infty}^{\infty} \int_{-\infty}^{\infty} I(x,y) e^{-i2\pi (ux + vy)} dx\,dy,
\label{eq:vc-z}
\ee
where $u$ and $v$ are the Fourier conjugates (spatial frequencies) of
the coordinates $x$ and $y$, respectively. The equation (\ref{eq:vc-z})
represents the van~Cittert--Zernike theorem (Born \&
Wolfe\cite{bo-wo}). The mutual coherence function $\Gamma (u,v)$
properly normalized (the process called ``calibration'' in practical
interferometry) is nothing but the cross-correlation function of the
signals detected by the two elements of the interferometer.

Equations (\ref{eq:wie-khi}) and (\ref{eq:vc-z}) form the foundation of
interferometry and aperture synthesis. Note, that the  mutual coherence
is a complex function. For convenience, it is usually represented as
$\Gamma (u,v) \equiv A(u,v) e^{i2\pi \phi(u,v)}$, where $A(u,v)$ and
$\phi (u,v)$ are the amplitude and phase, respectively. In principle,
the problem of reconstructing $I(x,y)$ requires a simple inverse
Fourier transformation of $\Gamma (u,v)$. However in practice, due to a
variety of reasons (telescope-specific instrumental and propagation
effects, to mention just a few), the measurement of $\Gamma (u,v)$ is
neither complete nor unbiased. The most serious problems surround
measurements of the phases of the interferometric response.

\section{ Closure relations in interferometry }

In the early days of radio interferometry, an ingenious method of
intensity interferometry had been introduced by Hanbury Brown \&
Twiss\cite{hb-tw} (see also Hanbury Brown\cite{hb74} and references
therein). In this method, the signals from two telescopes were first
square-law detected and filtered and then correlated. Problems with
measuring phases (as well as accurate synchronization of the receivers
at different telescopes) were thus eliminated. As a result, no proper
image could be produced, but the output of the correlator was
proportional to $<V_{1}V_{2}>^{2} \propto \Gamma(u,v)^{2}$. It was
certainly better than nothing as far as the structural properties of
the source were concerned. However, the method had a serious
disadvantage -- a lack of sensitivity. Signal handling by the intensity
interferometers requires amplification of the signals before
correlation, making the signal-to-noise detection threshold in the
incoming data considerably inferior to that of ``conventional''
interferometers. Nevertheless, this problem did not prevent successful
application of intensity interferometry in optical domain (also known
as speckle interferometry, see review by Bates\cite{bates}).

The intensity interferometry is perhaps worth mentioning here since it
resembles some techniques used in detection of particle production and
particle interferometry (see Cs\"{o}rg\H{o}\cite{csor} and references
therein). 

\begin{figure}
\centerline{
\psfig{file=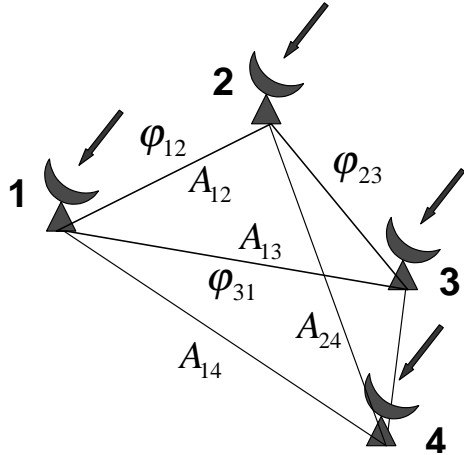,angle=-90,width=6.1cm,bbllx=138pt,bblly=231pt,bburx=420pt,bbury=514pt,clip=}
}
\caption{Radio telescopes 1,2,3 and 4 observe a distant celestial
source (which is located in the far zone; this allows us to consider
the electromagnetic waves shown by arrows as planar waves). Each of the
six two-telescope interferometers available in this scheme detect the
response represented by its amplitude $A_{ij}$ and phase $\phi_{ij}$
where $i,j=1,2,3,4$. Not all combinations of $A_{ij}$ and $\phi_{ij}$
are shown.}
\label{fig:closure}
\end{figure}

The next breakthrough in interferometry came with the introduction of
the so called ``closure'' relations. First, Jennison\cite{jen58}
conceived the idea of phase closure. It is based on the fact, that in
any interferometric triangle, the sum of the three interferometer
phases for a point-like source must be zero (see the interferometric
triangle ``1--2--3'' in Fig.~\ref{fig:closure}; the closure phase is
$\Phi_{123} \equiv \phi_{12} + \phi_{23}+ \phi_{31}$). If any
disturbance is imposed on the signal phase generated at an antenna in
such the triangle, it will be impressed on the interferometer phase on
two baselines linked to that antenna with opposite signs, thus leaving
the closure phase unchanged. For a non-point-like source, $\Phi_{123}
\neq 0$, but it remains invariant to any telescope-dependent
instrumental phase disturbances:
\be
\Phi_{123} \equiv \phi_{12} + \phi_{23}+ \phi_{31} = \psi_{12} + \psi_{23}+ \psi_{31},
\label{eq:cl-phas}
\ee
where $\psi_{ij}$ are measured (i.e. biased) interferometer phases. The
equation  (\ref{eq:cl-phas}) is a powerful constraint in recovering
true ``structural'' phases which represent the source structure via
equation (\ref{eq:vc-z}). Note, that baseline-dependent errors cannot
be cured by the closure phase relation (\ref{eq:cl-phas}).

Smith\cite{smi52} and Twiss et al.\cite{twiss} applied the idea of
closure relations to amplitudes. They showed that the value
$\mathcal{G} \equiv A_{12}A_{34}/(A_{13}A_{24})$ (dubbed ``closure
amplitude''; Fig.~\ref{fig:closure}) is independent of the gain (i.e.
amplitude) uncertainties.

\section { In place of conclusion }

This short presentation in no way is a complete description of the
broad and versatile technique of interferometry able to study
electromagnetic emission of celestial sources with unrivaled angular
resolution.  Instead, I hope, it can trigger further investigation of
possibly deep and potentially fruitful analogies between ``classical''
interferometry and multiparticle dynamics. It has to be noted that not
all possible parallels between the two disciplines have been mentioned
here. For example, the idea of correlation of photons produced in
W pair production has been suggested earlier by Chapovsky et
al.\cite{chap-kh}.

The following questions seem to be of interest:\\
1). What is the correspondence (if any) between measurable parameters in astrophysical interferometry and particle dynamics experiments? \\
2). If the parallel between intensity interferometry and some ``correlation'' detections in particle collisions is indeed meaningful, what is needed for the particle experiments to be treated as a ``conventional'' interferometer? The answer on this question could bring about applications of, e.g., closure relations, proven to be extremely valuable in astrophysical interferometry. 

In this publication I have chosen not to describe some recent
highlights of astrophysical interferometry reviewed in the verbal
version of this presentation. Those interested in the state-of-the-art
achievements of interferometers in various fields of astrophysics are
referred to the recent reviews in the Proceedings of the Symposium
``Galaxies and Their Constituents at the Highest Angular
Resolution''\cite{iau205} and references therein.

\section*{Acknowledgments}
I am grateful to the organizers of the Symposium and especially Wolfram
Kittel, Tam\'{a}s Cs\"{o}rg\H{o} and \v{S}arka Todorova for the
stimulating opportunity to look on the subject of interferometry from a
rather unusual (for the author) perspective. I thank Ian Avruch for
useful comments. I acknowledge the exchange programme of the Chinese
and Dutch Academies of Sciences.

\end{document}